\documentclass[preprint,superscriptaddress]{revtex4-1}
\usepackage{graphicx}
\usepackage{bm}
\usepackage{amsmath}
\usepackage{amsfonts}
\usepackage{amssymb}
\usepackage[utf8]{inputenc}
\newcommand {\be} {\begin{equation}}
\newcommand {\ba} {\begin{eqnarray}}
\newcommand {\ee} {\end{equation}}
\newcommand {\ea} {\end{eqnarray}}

\begin{document}

\title{Holographic description of elastic photon-proton and photon-photon scattering}

\author{Akira Watanabe}
\email{watanabe@ctgu.edu.cn (Corresponding author)}
\affiliation{College of Science, China Three Gorges University, Yichang 443002, People's Republic of China}
\affiliation{Center for Astronomy and Space Sciences, China Three Gorges University, Yichang 443002, People's Republic of China}
\author{Zabihullah Ahmadi}
\email{zabihullaha313@gmail.com}
\affiliation{College of Science, China Three Gorges University, Yichang 443002, People's Republic of China}
\author{Zhibo Liu}
\email{202107020021014@ctgu.edu.cn}
\affiliation{College of Science, China Three Gorges University, Yichang 443002, People's Republic of China}
\author{Wei Xie}
\email{xiewei@ctgu.edu.cn}
\affiliation{College of Science, China Three Gorges University, Yichang 443002, People's Republic of China}
\affiliation{Center for Astronomy and Space Sciences, China Three Gorges University, Yichang 443002, People's Republic of China}

\date{\today}

\begin{abstract}
We investigate the elastic photon-proton and photon-photon scattering in a holographic QCD model, focusing on the Regge regime.
Considering contributions of the Pomeron and Reggeon exchange, the total and differential cross sections are calculated.
While our model involves several parameters, by virtue of the universality of the Pomeron and Reggeon, for most of them the values determined in the preceding study on the proton-proton and proton-antiproton scattering can be employed.
Once the two adjustable parameters, the Pomeron-photon and Reggeon-photon coupling constant, are determined with the experimental data of the total cross sections, predicting the both cross sections in a wide kinematic region, from the GeV to TeV scale, becomes possible.
We show that the total cross section data can be well described within the model, and our predictions for the photon-proton differential cross section are consistent with the data.
\end{abstract}

\maketitle

\section{\label{sec:introduction}Introduction}
High energy hadron scattering has played an important role in investigating the partonic structure of the involved hadron for several decades.
Various scattering processes have been studied so far, but one of the simplest processes is the elastic hadron-hadron scattering, whose differential cross section is characterized by the center-of-mass energy $s$ and momentum transfer $t$.
Since this cross section is basically a nonperturbative physical quantity, it is difficult to theoretically predict it by the first principle calculation, although in some limited kinematic region, such as the high energy limit, it can be studied by the technique of perturbative quantum chromodynamics (QCD) and important results were obtained~\cite{Brodsky:1973kr, Matveev:1973ra, Lepage:1980fj}.
For the other kinematic region we need to rely on effective approaches, such as model calculations.
Moreover, it is especially difficult to theoretically analyze the scattering processes in the so-called Regge regime, in which the condition $s \gg t$ is satisfied, since the exchange of soft gluons may make the dominant contribution to the cross sections in the region.

The Regge theory~\cite{Veneziano:1968yb, COLLINS1971103, Collins:1977jy} has served as a useful approach to various high energy hadron scattering processes.
In this theory the scattering amplitude is obtained by taking into account the Pomeron and Reggeon exchange, which are interpreted as the multi-gluon and meson exchange, respectively.
This picture can also be realized in the framework of holographic QCD~\cite{Kruczenski:2003be, Son:2003et, Kruczenski:2003uq, Sakai:2004cn, Erlich:2005qh, Sakai:2005yt, DaRold:2005mxj, Brodsky:2006uqa}, which is an effective approach constructed based on the anti-de Sitter/conformal field theory (AdS/CFT) correspondence~\cite{Maldacena:1997re, Gubser:1998bc, Witten:1998qj}, and a lot of applications to the high energy scattering have been successfully made~\cite{Polchinski:2001tt, Polchinski:2002jw, Brower:2006ea, Hatta:2007he, Pire:2008zf, Domokos:2009hm, Domokos:2010ma, Marquet:2010sf, Watanabe:2012uc, Watanabe:2013spa, Watanabe:2015mia, Anderson:2016zon, Watanabe:2018owy, Xie:2019soz, Burikham:2019zbo, Watanabe:2019zny, Liu:2022out, Liu:2022zsa, Watanabe:2023rgp, Liu:2023tjr}.

In the preceding work~\cite{Liu:2022zsa}, the elastic proton-proton ($pp$) and proton-antiproton ($p \bar{p}$) scattering were investigated in holographic QCD, taking into account the Pomeron and Reggeon exchange contributions, and it was shown that the experimental data of the total and differential cross sections can be well described in a wide kinematic region.
We extend the previous work and investigate cross sections of the elastic photon-proton ($\gamma p$) and photon-photon ($\gamma \gamma$) scattering in this study.
Although a photon is a fundamental particle, an energetic photon can fluctuate into quark-antiquark pairs and behave like the lightest vector mesons, which is known as the vector meson dominance model developed by J. J. Sakurai~\cite{Sakurai:1960ju}.
Hence its cross sections can be hadronic, which enables us to apply the same framework as that used for the hadron-hadron cases.
Through this study we can further test the model presented in Ref.~\cite{Liu:2022zsa}, and improve our understanding of the photon structure.

In our model the Pomeron and Reggeon exchange are described by the Reggeized spin-2 glueball and vector meson propagator, respectively, and combining those with the Pomeron and Reggeon couplings at the vertices, the scattering amplitudes are obtained.
The model involves several parameters, but since the Pomeron and Reggeon trajectory related ones (intercepts and slopes) and the Pomeron-proton and Reggeon-proton coupling constants have been determined in the previous works~\cite{Xie:2019soz,Liu:2022zsa}, we can directly employ those values, due to the universality of the Pomeron and Reggeon.
The remaining two adjustable parameters, the Pomeron-photon and Reggeon-photon coupling constants, need to be determined, and this is made by numerical fitting with the experimental data of the $\gamma p$ and $\gamma \gamma$ total cross section.
We show that the both total cross section data can be well described within our model.
Once all the parameters are fixed, the differential cross sections can be predicted without any additional parameters, and it is presented that our predictions for the $\gamma p$ differential cross section are consistent with the data.
Besides, our predictions for the $\gamma \gamma$ differential cross section are also displayed.

This paper is organized as follows.
In Sec.~\ref{sec:holographic_model} we introduce how to describe the elastic $\gamma p$ and $\gamma \gamma$ scattering in holographic QCD, and derive the expressions of the cross sections.
The numerical results for the total and differential cross sections are presented in Sec.~\ref{sec:numerical_results}.
The conclusion of this work is given in Sec.~\ref{sec:summary}.

\section{\label{sec:holographic_model}Model setup}
In this section we introduce the holographic description of the elastic $\gamma p$ and $\gamma \gamma$ scattering in the Regge regime, considering contributions of the Pomeron and Reggeon exchange.
The lightest states of the Pomeron and Reggeon are assumed as the spin-2 glueball and vector meson, respectively, in this study.

\subsection{The Pomeron exchange}
For the proton-glueball-proton vertex, the energy momentum tensor matrix element $\langle p',s'|T^{\mu\nu}|~p,~s\rangle$ can be expanded with the three proton gravitational form factors as~\cite{Pagels:1966zza}
\begin{align}
\label{GFF_p}
\langle p',s' | T^{\mu\nu} | p,s\rangle = \bar{u}(p',s')\bigg[&A_{p}(t)\frac{\gamma^{\mu}P^{\nu}_p + \gamma^{\nu}P^{\mu}_p}{2}+B_{p}(t)\frac{ik_{p\rho}(P^{\mu}_p\sigma^{\nu\rho} + P^{\nu}_p\sigma^{\mu\rho})}{4m_{p}} \nonumber \\
~~+ &C_{p}(t)\frac{k^{\mu}_pk^{\nu}_p-\eta^{\mu\nu}k_p^2}{m_{p}}\bigg]u(p,s),
\end{align}
where $m_p$ is the proton mass, $k_p = p'_p-p_p$ is the momentum transfer, $t=-k_p^2$, and $P_p = (p'_p + p_p)/2 $.
When $|t|$ is quite small, the contribution of $B_p(t)$ term is negligible.
Since in this study we focus on the Regge regime, the final state can be regarded as the approximately same as the initial state ($p'\approx p$).

For the vector-glueball-vector vertex, the energy momentum tensor matrix element for the vector meson, which is denoted by $\rho$, is expressed as
\begin{align}
\label{GFF_rho}
&\langle \rho_n(p_1,\lambda_1)|T^{\mu\nu}(t)|\rho_n(p_2,\lambda_2)\rangle \nonumber \\
=& \epsilon^*_{2\alpha}\epsilon_{1\beta}\bigg[- A_\rho (t)\big(4k_\rho^{[\alpha}\eta^{\beta](\mu}p_\rho^{\nu)} + 2\eta^{\alpha\beta}p_\rho^\mu p_\rho^\nu\big) - 4 B_\rho (t) k_\rho^{[\alpha} \eta^{\beta](\mu}p_\rho^{\nu]} + \frac{1}{2} C_\rho (t) \eta^{\alpha\beta}(k_\rho^2 \eta^{\mu\nu} - k_\rho^\mu k_\rho^\nu ) \nonumber \\
&+ D_\rho (t) ( k^\alpha_\rho k^\beta_\rho \eta^{\mu\nu} - 2 k_\rho^{(\alpha}\eta^{\beta)(\mu}k_\rho^{\nu)} + k^2_\rho \eta^{\alpha(\mu}\eta^{\nu)\beta} ) + E_\rho (t) \frac{k_\rho^\alpha k_\rho^\beta}{m_\rho^2}p_\rho^\mu p_\rho^\nu + F_\rho (t) \frac{k_\rho^\alpha k_\rho^\beta}{m_\rho^2}(k_\rho^\mu k_\rho^\nu - k_\rho^2 \eta^{\mu\nu}) \bigg],
\end{align}
where $m_\rho$ is the $\rho$ meson mass, $p_\rho=(p'_\rho+p_\rho)/2$, $k_\rho=p'_\rho-p_\rho$, $A^{[\alpha}B^{\beta]} = (A^\alpha B^\beta-A^\beta B^\alpha)/2$, $A^{(\alpha}B^{\beta)} = (A^\alpha B^\beta+A^\beta B^\alpha)/2$, and $t=-k^2_\rho$.
$A_\rho(t)$, $B_\rho(t)$, $C_\rho(t)$, $D_\rho(t)$, $E_\rho(t)$, and $F_\rho(t)$ are the gravitational form factors.
It was shown in Ref.~\cite{Abidin:2008ku} that $B_\rho(t) = E_\rho(t) = 0$.
Furthermore, in the Regge limit the first term involving $A_\rho(t)$ in the right-hand side of the above equation only needs to be considered, and contributions of the other terms can be neglected.

To numerically evaluate the differential cross sections, whose results will be presented in the next section, we need to specify the functional forms of $A_{p}(t)$ and $A_\rho(t)$, which affect the $t$ dependence.
For these two gravitational form factors, in this study we utilize the results obtained with the bottom-up AdS/QCD models by the authors of Refs.~\cite{Abidin:2008ku, Abidin:2009hr}.
Their results involve a few parameters, but those can be determined with basic properties of the hadrons, such as the mass.
Hence the form factors we employ in this study do not bring any additional parameters into the present model.

Since we focus on the Regge regime, where $s \gg |t|$, in this study, the $t$-channel is the dominant channel.
The propagator of the spin-2 glueball is written as~\cite{Yamada:1982dx}
\begin{equation}
\label{propagator_1}
- \frac{id_{\alpha \beta \gamma \delta}(k)}{k^2 + m_g^2},
\end{equation}
where $m_g$ is the glueball mass and $d_{\alpha \beta \gamma \delta}(k)$ is given by
\begin{align}
\label{propagator_expand}
d_{\alpha\beta\gamma\delta}
= &\frac{1}{2}\big(\eta_{\alpha\gamma}\eta_{\beta\delta} + \eta_{\alpha\delta}\eta_{\beta\gamma}\big) - \frac{1}{2m_{g}^2}\big(k_{\alpha} k_{\delta}\eta_{\beta\gamma} + k_{\alpha}k_{\gamma}\eta_{\beta\delta} + k_{\beta}k_{\delta}\eta_{\alpha\gamma} + k_{\beta}k_{\gamma}\eta_{\alpha\delta}\big) \nonumber \\
&+ \frac{1}{24}\bigg[\left(\frac{k^2}{m_{g}^2}\right)^2 - 3\left(\frac{k^2}{m_{g}^2}\right) - 6\bigg]\eta_{\alpha\beta}\eta_{\gamma\delta} - \frac{k^2 - 3m_{g}^2}{6m_{g}^4}(k_{\alpha}k_{\beta}\eta_{\gamma\delta} + k_{\gamma}k_{\delta}\eta_{\alpha\beta}) \nonumber \\
&+ \frac{2k_{\alpha}k_{\beta}k_{\gamma} k_{\delta}}{3m_{g}^4}.
\end{align}
Only the first term in the right-hand side of this equation needs to be considered and contributions of the other terms can be neglected in the Regge limit.
Eq.~\eqref{propagator_1} can be rewritten as
\begin{equation}
D_{\alpha\beta\gamma\delta}^g=\frac{-i}{2(t - m_g^2)}\big(\eta_{\alpha\gamma}\eta_{\beta\delta} + \eta_{\alpha \delta}\eta_{\beta\gamma}\big).
\end{equation}
To include the higher spin states which lie on the leading Pomeron trajectory, the propagator needs to be Reggeized.
The factor $1/(t - m_g^2)$ is replaced with~\cite{Anderson:2016zon}
\begin{equation}
\frac{\alpha'_P}{2}e^{- \frac{i\pi\alpha_P(t)}{2}}\bigg(\frac{\alpha'_Ps}{2}\bigg)^{\alpha_P(t) - 2}\frac{\Gamma\big[3 - \frac{\chi}{2}\big]\Gamma\big[1-\frac{\alpha_P(t)}{2}\big]}{\Gamma\big[2-\frac{\chi}{2}+\frac{\alpha_P(t)}{2}\big]},
\end{equation}
where $\chi = \alpha_P(s)+\alpha_P(t)+\alpha_P(u)$, and $\alpha_P(x)=\alpha_P(0)+\alpha'_Px$ represents the Pomeron trajectory.
Hence $\chi$ can be rewritten as
\begin{equation}
\chi=3\alpha_P(0)+(s+t+u)\alpha'_P=3\alpha_P(0)+\Sigma_{i}m_i^2\alpha'_P,
\end{equation}
where $m_i$ represents masses of the four involved particles.

Taking into account the vector meson dominance, an energetic photon can fluctuate into quark-antiquark pairs and behave like the lightest vector mesons.
Since the Pomeron exchange gives the dominant contribution to the cross sections in the high energy region, $\gamma p$ ($\gamma \gamma$) scattering can be approximated as the $\rho$ meson-proton ($\rho$ meson-$\rho$ meson) scattering.
Therefore, the amplitudes of the $\gamma p$ and $\gamma \gamma$ scattering with the Pomeron exchange are expressed as
\begin{align}
\mathcal{A}_{\gamma p}^P&=\lambda_{g\gamma\gamma}\lambda_{gpp}2ip_\rho^\mu p_\rho^\nu\bar{u_4}\frac{iA_p(t)A_\rho(t)}{2}(\gamma^\delta p_p^\gamma+\gamma^\gamma p_p^\delta)\times D^g_{\mu\nu\delta\gamma} \nonumber
\\&=-\frac{i}{2}\lambda_{g\gamma\gamma}\lambda_{gpp}A_p(t)A_\rho(t)\times p_\rho^\mu p_\rho^\nu\bar{u_4}(\gamma^\delta p_p^\gamma+\gamma^\gamma p_p^\delta)u_2 \times(\eta_{\mu\delta}\eta_{\nu\gamma}+\eta_{\mu\gamma}\eta_{\nu\delta})\frac{\alpha'_P}{2}e^{-\frac{i\pi\alpha_P(t)}{2}}\nonumber
\\&~~~\times\bigg(\frac{\alpha'_Ps}{2}\bigg)^{\alpha_P(t)-2}~\frac{\Gamma\big[3-\frac{\chi_{\gamma p}}{2}\big]\Gamma\big[1-\frac{\alpha_P(t)}{2}\big]}{\Gamma\big[2-\frac{\chi_{\gamma p}}{2}+\frac{\alpha_P(t)}{2}\big]},
\end{align}
and
\begin{align}
\mathcal{A}_{\gamma\gamma}^P&=-2i\lambda_{g\gamma\gamma}^2A_\rho^2(t)p_\rho^\mu p_\rho^\nu p_\rho^\gamma p_\rho^\delta D_{\mu\nu\delta\gamma}^g\nonumber
\\&=-iA_\rho^2(t)\lambda_{g\gamma\gamma}^2s^2\frac{\alpha'_P}{2}e^{-\frac{i\pi\alpha_P(t)}{2}}\bigg(\frac{\alpha'_Ps}{2}\bigg)^{\alpha_P(t)-2}\frac{\Gamma\big[3-\frac{\chi_{\gamma\gamma}}{2}\big]\Gamma\big[1-\frac{\alpha_P(t)}{2}\big]}{\Gamma\big[2-\frac{\chi_{\gamma \gamma}}{2}+\frac{\alpha_P(t)}{2}\big]},
\end{align}
respectively, where $\chi_{\gamma p}=3\alpha_P(0)+2(m_p^2+m_\rho^2)$ and $\chi_{\gamma \gamma}=3\alpha_P(0)+4m_\rho^2$.
Then, the invariant amplitudes for the $\gamma p$ and $\gamma \gamma$ scattering can be obtained as
\begin{align}
\label{invariant1}
\bar{\mathcal{A}}_{\gamma p}^P&= - \sqrt{\frac{1}{2\times3}\Sigma_{\rm{spin}}|\mathcal{A}_{\gamma p}^P|^2}\nonumber
\\&= - \sqrt{\frac{1}{3}}\lambda_{g\gamma\gamma}\lambda_{gpp}A_p(t)A_\rho(t)s^2\times \frac{1}{t-m_g^2}\nonumber
\\&= - \frac{\sqrt{3}}{6}\lambda_{g\gamma\gamma}\lambda_{gpp}A_p(t)A_\rho(t)s^2\alpha'_Pe^{-\frac{i\pi\alpha_P(t)}{2}}\bigg(\frac{\alpha'_Ps}{2}\bigg)^{\alpha_P(t)-2}\frac{\Gamma\big[3-\frac{\chi_{\gamma p}}{2}\big]\Gamma\big[1-\frac{\alpha_P(t)}{2}\big]}{\Gamma\big[2-\frac{\chi_{\gamma p}}{2}+\frac{\alpha_P(t)}{2}\big]},
\end{align}
and
\begin{align}
\label{invariant2}
\bar{\mathcal{A}}_{\gamma \gamma}^P&= - \frac {1}{3}\lambda_{g\gamma\gamma}^2A_\rho^2(t)s^2\times\frac{1}{t-m_g^2}\nonumber
\\&= - \frac {1}{6}\lambda_{g\gamma\gamma}^2\alpha'_Pe^{-\frac{i\pi\alpha_P(t)}{2}}\bigg(\frac{\alpha'_Ps}{2}\bigg)^{\alpha_P(t)-2}\frac{\Gamma\big[3-\frac{\chi_{\gamma\gamma}}{2}\big]\Gamma\big[1-\frac{\alpha_P(t)}{2}\big]}{\Gamma\big[2-\frac{\chi_{\gamma\gamma}}{2}+\frac{\alpha_P(t)}{2}\big]},
\end{align}
respectively.

\subsection{The Reggeon exchange}
Here we consider the Reggeon exchange contribution to the $\gamma p$ and $\gamma \gamma$ scattering.
The propagator of the vector meson is expressed in the Regge limit as
\begin{equation}
D_{\mu\nu}^{v}(k) = -\frac{i}{t - m_v^2}\eta_{\mu\nu},
\end{equation}
where $m_v$ represents the vector meson mass.
To include the higher spin states lying on the Reggeon trajectory, the factor $1/(t-m_v^2)$ in the above equation needs to be replaced with~\cite{Anderson:2016zon}
\begin{equation}
\alpha'_{R}e^{-\frac{i\pi \alpha_{R}(t)}{2}}\sin\left[\frac{\pi\alpha_{R}(t)}{2}\right]\left(\alpha'_{R}s\right)^{\alpha_{R}(t) - 1}\Gamma[-\alpha_{R}(t)],
\end{equation}
where $\alpha_R(x)=\alpha_R(0)+\alpha'_Rx$ represents the Reggeon trajectory.
The photon-vector-photon and proton-vector-proton vertex are written as
\begin{equation}
\Gamma_{v\gamma\gamma}^{\mu}=-2i\lambda_{v\gamma\gamma}p_\gamma^\mu,
\end{equation}
and
\begin{equation}
\Gamma_{vpp}^{\nu} = - i \lambda_{vpp}\gamma^{\nu},
\end{equation}
respectively.
The amplitudes of the $\gamma p$ and $\gamma\gamma$ scattering with the Reggeon exchange are expressed as
\begin{align}
\mathcal{A}_{\gamma p}^R&=2\lambda_{v\gamma\gamma}\lambda_{vpp}p_\gamma^\mu\bar{u}_4\gamma^\nu u_2\eta_{\mu\nu}\times\frac{1}{t-m_v^2}\nonumber
\\&=2\lambda_{v\gamma\gamma}\lambda_{vpp}p_\gamma^\mu\bar{u}_4\gamma^\nu u_2\eta_{\mu\nu}\alpha'_{R}e^{-\frac{i\pi \alpha_{R}(t)}{2}}\sin\left[\frac{\pi\alpha_{R}(t)}{2}\right]\left(\alpha'_{R}s\right)^{\alpha_{R}(t)-1}\Gamma[-\alpha_{R}(t)],
\end{align}
and
\begin{align}
\mathcal{A}_{\gamma \gamma}^R&=4\lambda_{v\gamma\gamma}^2p_\gamma^\mu p_\gamma^\nu\eta_{\mu\nu}\times\frac{i}{t-m_v^2}\nonumber
\\&=2i\lambda_{v\gamma\gamma}^2s\alpha'_{R}e^{-\frac{i\pi \alpha_{R}(t)}{2}}\sin\left[\frac{\pi\alpha_{R}(t)}{2}\right]\left(\alpha'_{R}s\right)^{\alpha_{R}(t)-1}\Gamma[-\alpha_{R}(t)],
\end{align}
respectively.
Hence the invariant amplitudes for the $\gamma p$ and $\gamma \gamma$ scattering can be obtained as
\begin{align}
\label{invariant3}
\bar{\mathcal{A}}_{\gamma p}^R&=\sqrt{\frac{1}{1\times2}\Sigma_{spin}|\mathcal{A}_{\gamma p}^R|^2}\nonumber
\\&=2\lambda_{v\gamma\gamma}\lambda_{vpp}s\alpha'_{R}e^{-\frac{i\pi \alpha_{R}(t)}{2}}\sin\left[\frac{\pi\alpha_{R}(t)}{2}\right]\left(\alpha'_{R}s\right)^{\alpha_{R}(t)-1}\Gamma[-\alpha_{R}(t)],
\end{align}
and
\begin{align}
\label{invariant4}
\bar{\mathcal{A}}_{\gamma \gamma}^R&=\sqrt{\Sigma_{\rm{spin}}|\mathcal{A}_{\gamma\gamma}^R|^2}\nonumber
\\&=2\lambda_{v\gamma\gamma}^2s\alpha'_{R}e^{-\frac{i\pi \alpha_{R}(t)}{2}}\sin\left[\frac{\pi\alpha_{R}(t)}{2}\right]\left(\alpha'_{R}s\right)^{\alpha_{R}(t)-1}\Gamma[-\alpha_{R}(t)],
\end{align}
respectively.

Applying the optical theorem and combining those invariant amplitudes introduced above, the total cross sections of the $\gamma p$ and $\gamma\gamma$ scattering with both the Pomeron and Reggeon exchange are given by
\begin{align}
\sigma_{\rm{tot}}^{\gamma p(\gamma\gamma)}=\frac{1}{s} ~{\rm Im}(\bar{\mathcal{A}}^P_{\gamma p(\gamma\gamma)}+\bar{\mathcal{A}}^R_{\gamma p(\gamma\gamma)}).
\end{align}
The differential cross sections are given by
\begin{equation}
\label{dcs final1}
\frac{d\sigma^{\gamma p(\gamma\gamma)}}{dt}=\frac{1}{16\pi s^2}|\bar{\mathcal{A}}^P_{\gamma p(\gamma\gamma)}+\bar{\mathcal{A}}^R_{\gamma p(\gamma\gamma)}|^2.
\end{equation}
These cross sections are numerically evaluated, and the results are presented in the next section.

\section{\label{sec:numerical_results}Numerical results}
Our analytical results for the cross sections involve eight parameters in total, and here we explain how to fix them.
Considering the universality of the Pomeron and Reggeon, for the trajectory related parameters (intercepts and slopes) we can utilize the results obtained in the previous works.
Moreover, since the Pomeron-proton and Reggeon-proton coupling constant were also determined in the previous works, we can employ those results.
For the Pomeron related parameters $\{\alpha_P(0),~\alpha'_P,~\lambda_{gpp} \}$, we utilize the results obtained in Ref.~\cite{Xie:2019soz}, in which the elastic $pp$ and $p \bar{p}$ scattering at high energies were investigated by only considering the Pomeron exchange contribution.
For the Reggeon related parameters $\{\alpha_R(0), ~\alpha'_R,~\lambda_{vpp} \}$, we utilize the results obtained in Ref.~\cite{Liu:2022zsa}, in which contributions of both the Pomeron and Reggeon exchange were considered to investigate the $pp$ and $p \bar{p}$ scattering in the medium energy region.
The parameter values used in this study are summarized in Table~\ref{table}.
\begin{table}[tb]
\centering
\caption{Parameter values.}
\begin{tabular}{l l l}
\hline
Parameter&~~~~Value&~~~~Source\\                
\hline
$\alpha_P(0)$ &~~~~1.086 &~~~~fit to $pp(p \bar{p})$ data at high energies~\cite{Xie:2019soz} \\
$\alpha'_P$ &~~~~0.377~$\rm{GeV}^{-2}$ &~~~~fit to $pp(p \bar{p})$ data at high energies~\cite{Xie:2019soz} \\
$\lambda_{gpp}$ &~~~~9.699~$\rm{GeV}^{-1}$ &~~~~fit to $pp(p \bar{p})$ data at high energies~\cite{Xie:2019soz} \\
$\alpha_R(0)$ &~~~~0.444 &~~~~fit to $pp(p \bar{p})$ data at medium energies~\cite{Liu:2022zsa} \\
$\alpha'_R$ &~~~~0.925$~\rm{GeV}^{-2}$ &~~~~fit to $pp(p \bar{p})$ data at medium energies~\cite{Liu:2022zsa} \\
$\lambda_{vpp}$ &~~~~7.742~ &~~~~fit to $pp(p \bar{p})$ data at medium energies~\cite{Liu:2022zsa} \\
$\lambda_{g\gamma\gamma}$ &~~~~0.04078~$\pm~7.6\times 10^{-7}~\rm{GeV}^{-1}$ &~~~~this work \\
$\lambda_{v\gamma\gamma}$ &~~~~0.03447~$\pm~4.9\times10^{-6}$ &~~~~this work \\
\hline
\end{tabular}
\label{table}
\end{table}

\begin{figure}[tb]
\centering
\includegraphics[width=0.6\textwidth]{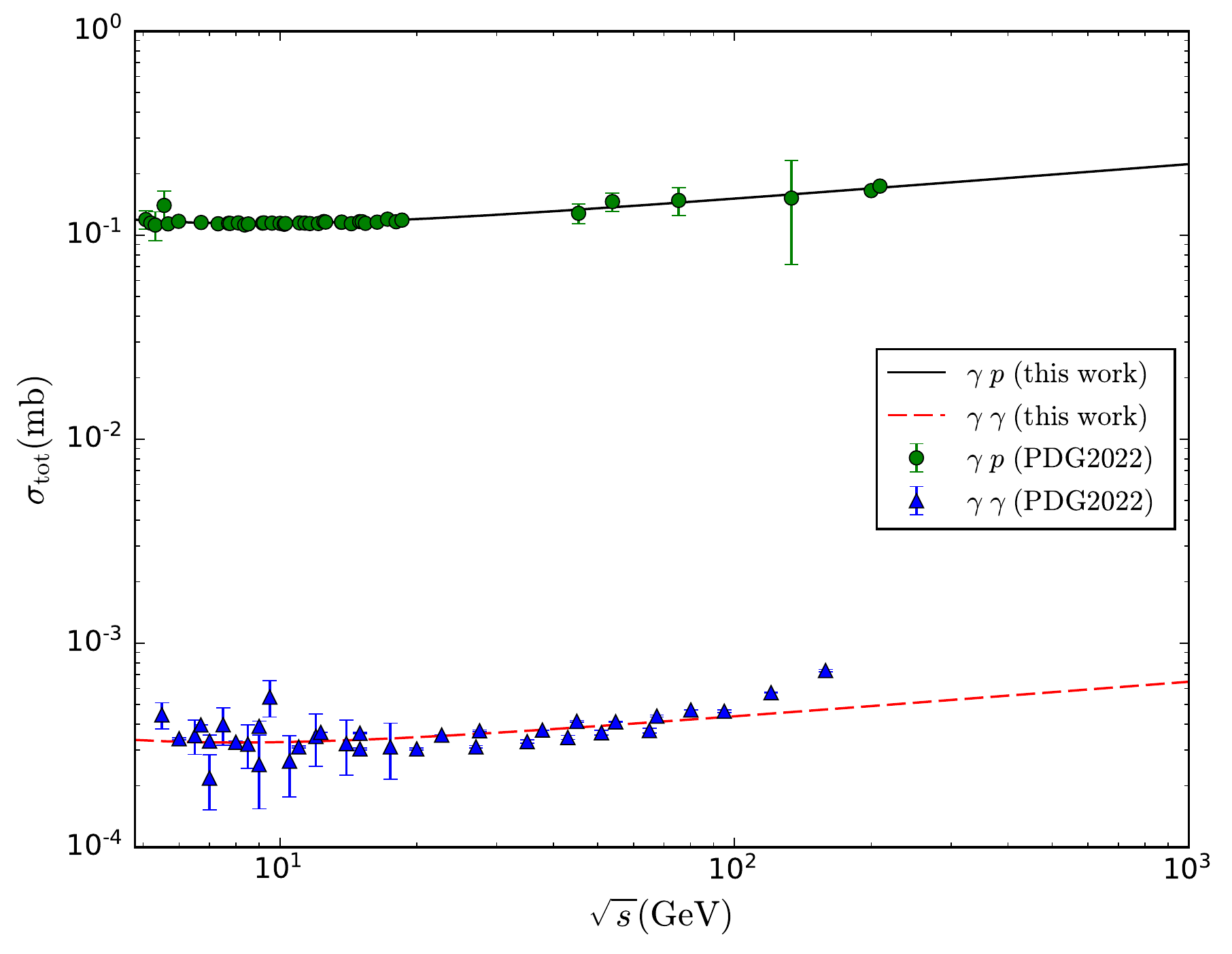}
\caption{
The total cross section as a function of $\sqrt{s}$.
The solid and dashed curve represent our calculations for the $\gamma p$ and $\gamma \gamma$ scattering, respectively.
The experimental data are taken from %
Ref.~\cite{ParticleDataGroup:2022pth}.
}
\label{TCS}
\end{figure}

There are two adjustable parameters, $\lambda_{g\gamma\gamma}$ and $\lambda_{v\gamma\gamma}$, which are the Pomeron-photon and Reggeon-photon coupling constant, respectively.
We determine these by numerical fitting, using the experimental data of the $\gamma p$ and $\gamma \gamma$ total cross section at $\sqrt{s} \geq 5$~GeV, which are summarized by the Particle Data Group (PDG) in 2022~\cite{ParticleDataGroup:2022pth}, and the MINUIT package~\cite{James:1975dr}.
The obtained parameter values are shown in Table~\ref{table}, and the resulting total cross sections are displayed in Fig.~\ref{TCS}.
It is found from the figure that the experimental data are well described with our model in the considered kinematic region.

\begin{figure}[tb]
\centering
\includegraphics[width=1\textwidth]{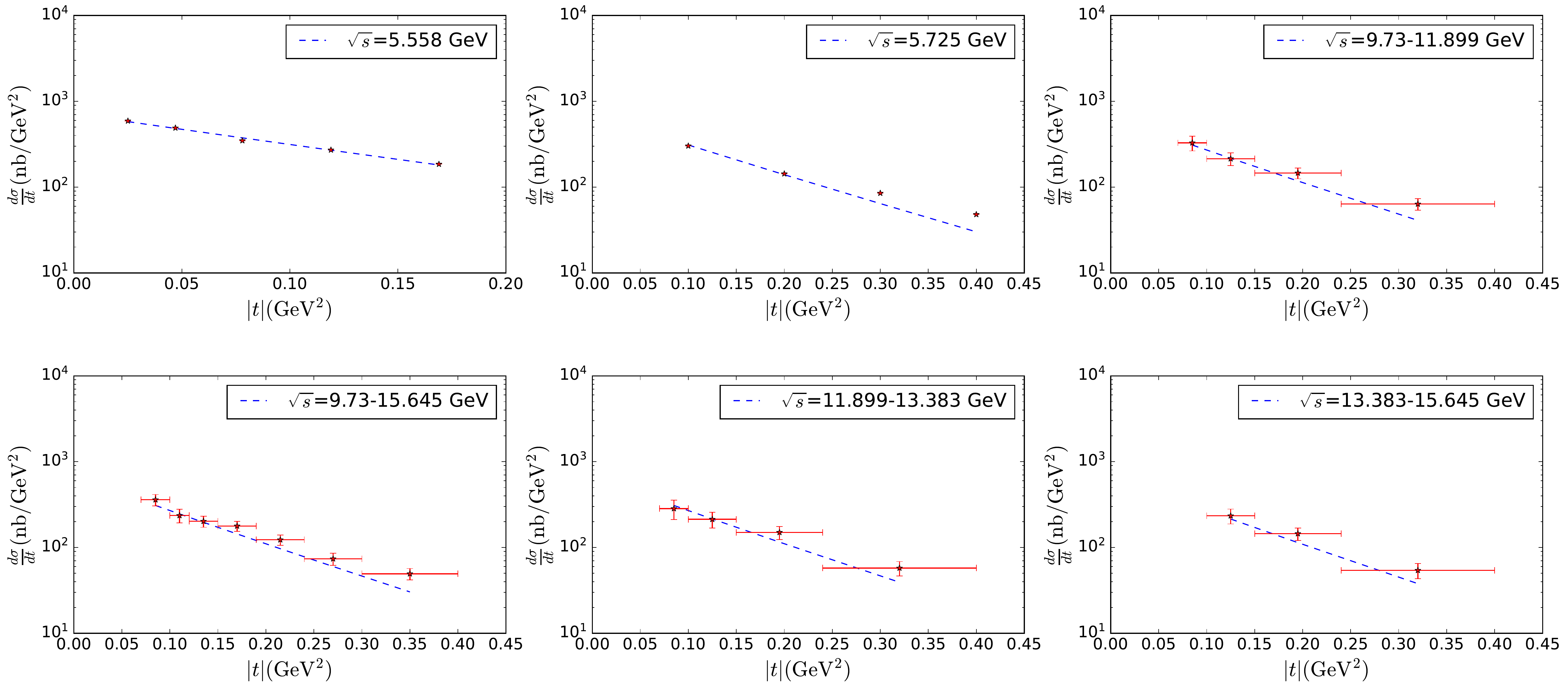}
\caption{
The differential cross section of the $\gamma p$ scattering as a function of $|t|$ for various $\sqrt{s}$.
The dashed curves represent our predictions.
The experimental data are taken from Refs.~\cite{Criegee:1977ue, Anderson:1970wz, Breakstone:1981wm, Buschhorn:1971nw}.
}
\label{DCS}
\end{figure}

\begin{figure}[tb]
\centering
\includegraphics[width=0.9\textwidth]{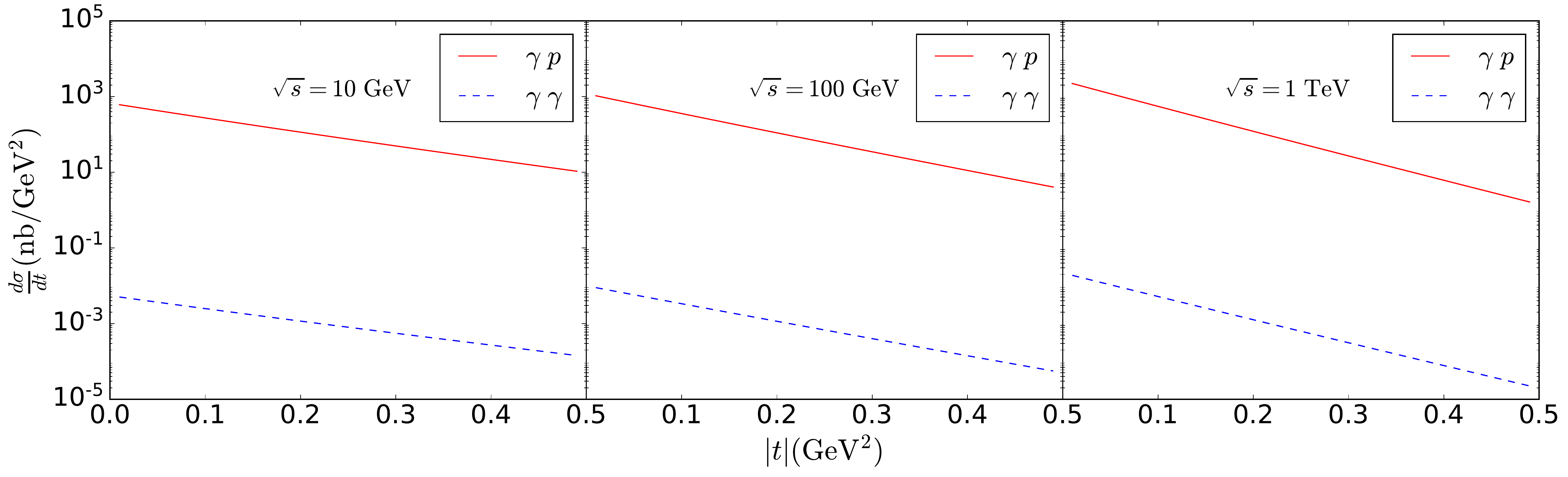}
\caption{
The differential cross section as a function of $|t|$ for $\sqrt{s} =$ 10~GeV, 100~GeV, and 1~TeV.
The solid and dashed curves represent our predictions for the $\gamma p$ and $\gamma \gamma$ scattering, respectively.
}
\label{DCS_predict}
\end{figure}

Once the two adjustable parameters are determined with the total cross section data, we can predict the differential cross sections.
To avoid the influence of the Coulomb interaction in the very small $|t|$ region~\cite{Amos:1985wx, UA4:1987gcp} and focus on the Regge regime, we limit the $|t|$ range to $0.01 \leq |t| \leq 0.45$~GeV$^2$.
We display the comparisons between our predictions and the experimental data~\cite{Criegee:1977ue, Anderson:1970wz, Breakstone:1981wm, Buschhorn:1971nw} for the $\gamma p$ differential cross section at $\sqrt{s} \geq 5$~GeV in Fig.~\ref{DCS}.
It is seen that our predictions are consistent with the data, although the available data concentrate in somewhat narrow $\sqrt{s}$ range.
In the panel showing the results at $\sqrt{s} = 5.725$~GeV, there are some deviations between our predictions and the data, which may be due to the relatively larger $t/s$ value, compared to the other results.
Since currently there is no other available data for the differential cross sections, we display our predictions for both the $\gamma p$ and $\gamma \gamma$ scattering in Fig.~\ref{DCS_predict}.
%

\section{\label{sec:summary}Conclusion}
We have investigated the elastic $\gamma p$ and $\gamma \gamma$ scattering in the Regge regime, considering the Pomeron and Reggeon exchange in a holographic QCD model.
Combining the Reggeized spin-2 glueball and vector meson propagator, which describe the Pomeron and Reggeon exchange, respectively, and the Pomeron and Reggeon couplings at the vertices, the scattering amplitudes are derived and expressions for the total and differential cross sections are obtained.
While those expressions involve several parameters, for most of them we can utilize the values determined in the previous works on the $pp$ and $p \bar{p}$ scattering, and there are only two adjustable parameters which need to be determined in this study.

We have determined the two parameters with the experimental data of the total cross sections, and shown that the data for both the $\gamma p$ and $\gamma \gamma$ scattering can be well described within the present model.
Then, we have presented our predictions for the differential cross sections, which can be calculated without any additional parameters.
Our results for the $\gamma p$ scattering are consistent with the data, which implies that the scattering amplitude is correctly obtained.

Although the vector meson dominance is known as an experimental fact, it is interesting that the photon involved processes can also be well described with the present framework.
Especially for the differential cross sections, the currently available data are quite limited.
It is expected that more data will help to further test the present model in the future, and to deepen our understanding of the photon nature, which is one of the most fundamental issues in high energy physics.

\section*{Acknowledgments}
The work of A.W. was supported by the start-up funding from China Three Gorges University (CTGU).
A.W. is also grateful for the support from the Chutian Scholar Program of Hubei Province.
The work of W.X. was supported by CTGU Teaching and Research Grant No. J2022050.

\bibliography{references}

\end{document}